\documentclass[12pt,preprint]{aastex}

\newcommand{\grad}{\mbox{$\,^\circ$}}

\shorttitle{Kinematics of ISOHDFS galaxies} \shortauthors{Rigopoulou
et al.}

\begin{document}


\title{Kinematics of Galaxies in the Hubble Deep Field South:
Discovery of a Very Massive Spiral at z=0.6 \footnote{Based on 
observations collected at the European Southern
Observatory, Chile, ESO Nos 65.O-0612, 67.A-0518}}


\author{D. Rigopoulou\altaffilmark{1}, A. Franceschini\altaffilmark{2},
H. Aussel\altaffilmark{3}, R. Genzel\altaffilmark{1}, 
N. Thatte\altaffilmark{1},
C.J. Cesarsky\altaffilmark{4}}
\altaffiltext{1}{Max-Planck-Institut f\"ur
extraterrestrische Physik,  Postfach 1312, 85741 Garching, Germany}
\altaffiltext{2}{Dipartimento di Astronomia, Universita' di Padova,  
Vicolo Osservatorio 5 I-35122, Padova, Italy}
\altaffiltext{3}{Institute for Astronomy, 2680 Woodlawn Drive, 
Honolulu, Hawaii, 96822, USA}
\altaffiltext{4}{European Southern Observatory,
Karl-Schwarzschild-str. 2,  85740 Garching, Germany}


\begin{abstract}

We report the first results from a study of the internal kinematics,
based on spatially resolved H$_{\alpha}$ velocity profiles, of
three galaxies at redshift z$\sim$0.6 and one at redshift z$\sim$0.8, 
detected by ISOCAM in the Hubble Deep Field South. The kinematics are derived
from high resolution near-infrared VLT spectroscopy. 
One of the galaxies is a massive spiral which possesses 
a very large rotational velocity of 460 km s$^{-1}$ and contains a mass
of 10$^{12}$ M$_{\odot}$ (within 20 kpc),
significantly higher than the dynamical masses measured in most other local
and high redshift spirals. Two of the galaxies comprise a counter-rotating 
interacting system, while the fourth is also a large spiral.
The observed galaxies are representative examples of the morphologies 
encountered among ISOCAM galaxies.
The mass-to-light (M$/$L$_{bol}$) ratios of ISOCAM galaxies lie between
those of local luminous IR galaxies and massive spirals.
We measure an offset of  1.6$\pm$0.3 mag in the rest frame 
B-band and of 0.7$\pm$0.3 mag in the rest frame I-band when 
we compare the four 
ISOCAM galaxies to the local Tully-Fisher B and I-band relations. 
We conclude that the large IR luminosity of the ISOCAM population results from
a combination of large mass and efficient triggering of star formation.
Since ISOCAM galaxies contribute
significantly to the Cosmic Infrared Background our results 
imply that a relatively small number of very massive and IR luminous objects
contribute significantly to the IR background and star formation activity near
z$\sim$0.7.

\end{abstract}


\section{Introduction}

Most studies of high redshift galaxies have targeted their 
statistical properties such as galaxy counts, colors and redshift 
distributions (e.g. Koo \& Kron~1992; Lilly et al.~1995).
The information obtained this way has been used to construct 
the luminosity
function which when compared with various models can give us clues about the
evolution of galaxies at earlier epochs (e.g. Gronwall \& Koo~1995). 
The advent of 10m class telescopes
combined with advances in detector technology offers the possibility of 
a more detailed investigation of the high redshift population aimed at better
understanding their physical properties and internal kinematics, the latter 
being ultimately related to the question of their evolution. 
Such a detailed study of the internal kinematics of high-redshift systems 
has the advantage of studying the evolution
of galaxies on a galaxy by galaxy basis, in contrast to redshift 
surveys that study a large sample of similar but not exactly identical galaxies
and derive ``mean'' properties that are then applied to the entire sample.

However, measuring the evolution of mass$/$luminosity of high-redshift 
galaxies directly is a challenging task. 
The Faber-Jackson relation for spheroid-dominated systems 
(Faber \& Jackson~1976) and the Tully-Fisher
relation for spiral galaxies  (Tully \& Fisher~1977, TF) are important
tools for investigating galaxy kinematics. By comparing high-redshift
galaxies with morphologically similar local galaxies we can study fundamental
properties of the high-z systems, investigate the processes by 
which disk-formation occurs (e.g. Navarro \& Steinmetz 2000; Mo \& Mao 2000) 
and eventually understand galaxy formation. Internal kinematic information 
is crucial since it provides a direct estimate of the masses of galaxies 
and allows galaxy evolution to be traced directly by mass as distinct from
light. 
To date, datasets on kinematics of high-redshift galaxies contain few galaxies
and the results have been somewhat discrepant. For instance,
Rix et al.~(1997), Simard \& Pritchet (1998) find significant luminosity
evolution for their low to intermediate (0.2$<$z$<$0.4) redshift galaxies 
while Vogt et al.~(1996; 1997) find evidence for only modest brightening
for their (0.1$<$z$<$1) sample. 
Apart from measuring offsets from the local TF, 
examining differences in the slope of the TF is also crucial 
(e.g. Barton et al.~2001). For 
instance Ziegler et al.~(2002) report that the TF slope changes with redshift.
An increase in the TF slope with redder passbands going from L$\propto$v$^{3}$
in the optical to L$\propto$v$^{4}$ in the near infrared has also been
found (e.g. Verheijen~2001).

The discovery by COBE of an extragalactic infrared background with an intensity
at least as large as that of the optical background suggests that a significant
fraction of the star formation in the Universe takes place in dust enshrouded
regions and is thus missed by optical surveys.
With the advent of the Infrared Space Observatory (ISO,  Kessler et
al.~1996) deep mid--IR surveys for distant galaxies have
been successfully carried out for the first time.  
Operating in the 5 -- 18 $\mu$m band
sensitive to warm dust and PAH$/$UIB emission features, ISOCAM on
board ISO was more than 1000 times more sensitive than IRAS and  thus
had the potential to study infrared bright galaxies at redshifts
beyond 0.5.  A number of cosmological surveys have been performed with
ISOCAM especially in the LW3 filter (12 -- 18$\mu$m) resulting in the 
detection of galaxies down to flux density limits of 100 $\mu$Jy.
We are engaged in a followup effort to investigate the nature of
galaxies discovered through deep ISOCAM surveys. 
This new population of galaxies is at z$\sim$1 and 
shows strong cosmological evolution (Elbaz et al.~1999, 
Franceschini et al.~2001). 
In Rigopoulou et al.~(2000), \citetext{hereafter Paper I} we 
addressed the issue of the energetics by studying the  
properties of a sample of galaxies detected by ISOCAM 
observations of the Hubble Deep Field South (HDFS, hereafter ISOHDFS galaxies).
Based on our low resolution near infrared (rest-frame R-band) spectra we 
concluded that ISOHDFS galaxies are powerful dusty starbursts, 
consistent with their large mid-IR luminosities. These 
observations indicate that this population traces an important phase of galaxy
evolution. The next and 
most challenging step is to investigate their mass content and to compare it 
with the ongoing rate of star formation quantified in Paper I.

In this paper we present our results on spatially resolved
rotation curves of four ISOHDFS galaxies selected from the sample 
of ISOCAM detections in the Hubble Deep Field South. 
We compare their properties to those of local galaxies and investigate 
the offset from the local TF.
We find that one of the four galaxies studied is
a {\em \bf{ massive spiral with an extraordinarily high mass of
10$^{12}$\ M$_{\odot}$}}.


\section{Sample Selection}

The ISOCAM observations of the Hubble Deep Field South 
(HDFS, Oliver et al.~2002), Aussel et al.~in preparation) were carried 
out with two broad--band filters, LW2 ( 5--8 $\mu$m, 
$\lambda_{eff}$ = 6.75 $\mu$m) and LW3 (12--18 $\mu$m, 
$\lambda_{eff}$ = 15 $\mu$m). The ISOCAM observations were 
centered on the HST--WFPC2 field extending over 25 sq. arcmins,
details of the observations can be found
in Oliver et al.~ (2002). In the following we will only consider the 
LW3 15$\mu$m sample. 

Oliver et al.~(2002) and Aussel et al.~(in preparation) analyzed the data using two
independent methods. Aussel et al.~used the PRETI method 
(Starck et al.~1999)
which is based on a wavelet analysis of the combined spatial and temporal 
observable space.
The LW3 observations resulted in the detection of 
63 sources reaching down to a limiting flux of $\sim$100 $\mu$Jy (from
the analysis of Aussel et al.~2002). For our followup studies we selected
objects from the Aussel et al.~analysis based on the following criteria:
a) a reliable LW3 detection, b) H$_{\alpha}$ in the wavelength range 
of the ISAAC spectrometer (Z, SZ, J, H and K bands) and, c) a secure 
counterpart in the I band image and$/$or a
counterpart in the K band image (EIS Deep, daCosta et al.~1998). We
did not apply any selection based on colors. Based on these criteria our
ISOCAM HDFS (hereafter ISOHDFS) sample contains about 25 galaxies 
with 15 $\mu$m flux densities 
ranging between 100--800 $\mu$Jy and provides a fair sampling of the
strongly evolving ISOCAM population near the peak of the differential
source counts (Elbaz et al.~1999). Near-infrared (rest-frame R-band)
and optical (rest-frame B-band) low resolution spectra for our ISOHDFS sample
have been presented in Paper I and in Franceschini et al.~(2002).

From our ISOHDFS sample we selected randomly a few candidate 
objects for
the high resolution studies presented here. In Table 1 we present the
target's coordinates and 
the available ground based photometric data (AB magnitudes) from the ESO--EIS
Deep Survey (daCosta et al.~1998). The coordinates of the galaxies
have been taken from Franceschini et al.~(2002). 
In Table 2 we present morphological type (from the work of Lanzetta et al.~, 
see footnote number 7) 
redshifts and H$_{\alpha}$ flux values (from Rigopoulou et al.~2000), LW3 fluxes and
extinction estimates (from Franceschini et al.~2002) 
and target sizes as measured from the HDF-S WFPC2 image and$/$or the 
EIS-K band image. {\em The properties
and morphological types of the 
candidate galaxies are representative of the properties of the entire 
sample of ISOCAM galaxies (for a discussion of the morphologies of ISOCAM
galaxies see section 7).}

\section{Data}

\subsection{Acquisition and Reduction}

We took the galaxy spectra presented here during 2000, August 20--24, and
2001,  August 29--30 using the near-infrared spectrometer ISAAC
(Moorwood~1998) on  the ANTU--ESO telescope (formerly UT1), on
Paranal, Chile.  For the observations we used the medium resolution
grating  R$_{s} \sim $ 5000 and a 0.6$^{\arcsec}$$\times$2$^{\arcmin}$
long slit. The orientation of the slit was along the major axis of
ISOHDFS 27 (PA: 49$\grad$) and along the two nuclei in  ISOHDFS 25 (PA:
65.5$\grad$) (as determined from HST WFPC2 images for ISOHDFS 27 and
near-IR images for ISOHDFS 25).  For the observations of ISOHDFS 55 we
positioned the slit at 52$\grad$ so as to include the bright spot SE
of the nucleus (the rotation curve has been corrected for the
difference between the PA of the slit and the major axis).  The total
on-source integration time was 10800 sec for each galaxy with
individual exposures being 300 sec long.  Observations were done by
nodding the telescope $\pm$ 20$^{\arcsec}$  along the slit  to
facilitate sky subtraction.  Sky conditions were excellent throughout
the acquisition of the ISOHDFS 27 and  ISOHDFS 25  spectra,  with
seeing values in the range 0.25$^{\arcsec}$--0.6$^{\arcsec}$  while
the ISOHDFS 55 spectrum was acquired through less optimal conditions
(seeing $\sim$0.8$^{\arcsec}$).  Following the acquisition of the
galaxy spectra observations we observed spectroscopic standard stars
in order to flux calibrate the galaxy spectra.

We reduced the ISAAC data using standard applications from the
IRAF\footnote{IRAF is distributed by the National Optical Astronomy
Observatories, which are operated by the Association of Universities
for Research in Astronomy, Inc. under the cooperative agreement with
the National Science Foundation.} package.  We took spectra in
nod-along-the slit pairs.  To remove sky background (including OH
lines) we subtracted the two frames  of a pair and flat-fielded the
result. We wavelength calibrated our frames using neon-argon arc-lamp
frames. We then removed hot$/$bad pixels  from the sky, star and
science frames and corrected for the curvature of the slit. We
transformed each exposure onto a linear wavelength grid with the
dispersion  axis parallel to detector rows.  The sky-subtracted,
co-added galaxy spectra were further divided by the spectrum of the
spectroscopic standard star in order to correct for atmospheric
transmission. Finally we extracted galaxy spectra using  {\em apall}
from the {\em twodspec} package of IRAF.

\subsection{Light profiles}

The galaxy luminosity profiles were measured using ELLIPSE in the
STSDAS$/$ISOPHOTE package which fits isophotal ellipses to the galaxy
luminosity profiles. The fits for ISOHDFS 27 and ISOHDFS 55 were made
using the I$_{814}$ image and are shown in Figure 1. The dashed lines
are exponential profile fits, the disk-scale lengths are reported in
Table 3. 
Since ISOHDFS 25 is outside of HDFS WFPC2 and flanking
fields we did not  have any high resolution imaging data.

\subsection{Kinematic Modelling}

To derive the rotation velocities of the galaxies as a function of
position, we fit Gaussian profiles to the strongest
emission line H$_{\alpha}$.  We performed Gaussian fits using the
application NGAUSSFIT within the IRAF$/$STSDAS package.
The background was removed prior to applying the Gaussian fits.   The
widths were constrained to a physically meaningful value (1--20\.A).
To increase the signal-to-noise ratio (S$/$N) of ISOHDFS 55 we
smoothed the  raw data by a boxcar filter of 0.65$\arcsec$ along the spatial
axis before the  measurement.

Deteriming the true rotational velocity V$_{ rot}$ of high-z galaxies
is,  unlike the case for local spirals, far from trivial. For one, the
sizes of high-z galaxies are typically similar to the slit widths
($\sim$ 0.6$\arcsec$--1$\arcsec$)  so the resultant spectrum 
is an integration over
the galaxy's intrinsic  velocity fields. Various methods have so far
been used to measure  rotational velocities in distant galaxies
including seeing modelling of  the emission lines (e.g. Vogt et
al. 1996, 1997), a maximum-likelihood  technique based on properties of
local galaxies (e.g. Rix et al.~1997)  or a synthetic rotation curve
(e.g Simard and Pritchet 1999, Ziegler et al.~2002).

To interpret our data we constructed a simple synthetic rotation curve
(RC). We used an exponential disk to model the surface brightness
distribution. The equation:\\
\begin{center}
$S(r) = S_{o} e ^{-r/r_{\small d}}$\\
\end{center}

where S$_{o}$ and r$_{\small d}$ are the central surface brightness
and scale length of the disk, describes the surface brightness as a
function of radius. In reality the surface brightness profiles of
galaxies are  influenced by dust absorption, however, 
we do not consider these effects in the present
model. Although one of the galaxies
studied here,  the spiral ISOHDFS 27 most likely includes a bulge, our
model does not include a bulge component since  (a) our observations
indicate that the emission lines originate in the disk, (b) bulge-disk
deconvolution showed that the contribution of the bulge is much
smaller than the disk and (c) we prefer to use as few model parameters
as possible.  We assumed that the model velocities rise
linearly with radius up to one disk scale length and become flat
(constant velocity)  at larger radii (Persic \& Salucci~1991, PS).
Simard \& Pritchet (1998) who used a slightly different velocity field
(a step function ``flat'' rotation curve) in their model, found no
difference between their velocity field and the PS velocity field.  We
computed the radius at which the intrinsic velocity flattens  from the
disk  scale lengths r$_{\small d}$ measured in the F814W band keeping
in mind that the scale length of the younger stellar population is
larger than that of the older stellar population, (Ryder \& Dopita~
1994).  This intrinsic one-dimensional velocity field was then  used
to generate a 2D velocity field taking into account inclination and
position angle. We measured inclination angles from HST$/$WFPC2 F814W
images for the two galaxies ISOHDFS 27 and 55 and from the EIS K-band
image for ISOHDFS 25. The errors on the measured inclination angles
are small of the order of $\pm$2$^{\grad}$ for ISOHDFS 27 and 55 and
$\pm$5$^{\grad}$ for ISOHDFS 25. 
We then convolved our disk
with the appropriate Gaussian to account for the seeing and extracted
a simulated Rotation Curve (RC) from the 2D velocity field by integrating 
over the
slit aperture.  The only free parameter in our model is the true
rotation velocity  V$_{ rot}$. We varied V$_{ rot}$ until the
extracted RC best matched the  observed one. We then adopted this
V$_{rot}$ value as the true rotational velocity of the galaxy.
Errors in V$_{rot}$ were estimated by varying the inclination and position
angle of each galaxy by a small amount ($\pm$12$^{\grad}$) and using the 
extreme values.

\section{Results}

Table 2 gives the fitting data for the galaxies studied
here. Throughout this work we have assumed H$_{\small 0}$ = 70 km
s$^{-1}$ Mpc$^{-1}$,  $\Omega_{M}$ =0.3, $\Omega_{\Lambda}$=0.7 .  
The galaxies have been corrected
for extinction using the Balmer  decrement (Franceschini et al.~ 2002),
and internal extinction following  the prescription by Tully \&
Fouqu\'e (1985).

The extracted spectra of the targets appear in Figures 2--4.  For each
galaxy we show the HST$/$WFPC2 (F814 filter) and$/$or  K-band ground
based image (for ISOHDFS 25) with the slit orientation drawn on it,
the sky-subtracted 2-D spectra, the (observed) velocity profile and
finally the observed rotation curves  together with the model fits.
Two of the galaxies studied here, ISOHDFS 27, and ISOHDFS 55 are large
Sb type spirals\footnote{from the classification of K. Lanzetta and
collaborators which can be found in
www.ess.sunysb.edu$/$astro$/$hdfs$/$wfpc2).}  while ISOHDFS 25 is an
interacting galaxy comprising two (as derived from the analysis)
counterotating galaxies. For the two Sb galaxies we performed fits to
the light profiles (see section 3.2) which allowed us to derive
bulge and disk parameters. The bulge-to-disk ratio is 0.07 and 0.02 for
ISOHDFS 27 and ISOHDFS 55, respectively. 
The positions of the strong nebular lines
H$_{\alpha}$ and [NII] are labelled.  As we will discuss in section 7
the galaxies studied here are representative of the morphologies found
among the samples of ISOCAM galaxies.  We next comment on the
individual sources.

{\bf ISOHDFS 27}:  The H$_{\alpha}$ emission is clearly spatially
extended over $\sim$4$^{\arcsec}$ (FWHM) which at the distance of
ISOHDFS 27 corresponds to about $\sim$ 25 kpc (z=0.58,
1$^{\arcsec}$=5.92 kpc).   Two lobes are detected on each side of the
H$_{\alpha}$ continuum (same structure  is seen in [NII]) providing
evidence that the emission originates  in a rotating disk. That the
H$_{\alpha}$ line emission originates in a disk is also supported by
the characteristic double-horn shape seen in the line velocity
profile.  The H$_{\alpha}/$[NII] line ratio is about 3 which is typical of
HII regions.  This line ratio combined with the extended  morphology
of the H$_{\alpha}$ line on the one hand and the
LW3(15$\mu$m)$/$LW2(7$\mu$m) flux ratio on the other,  rules out the
presence of a central AGN as the  prime ionizing source (as already
suggested in Paper I).  An interesting aspect of the spectrum of
ISOHDFS 27 is its velocity width.  We have measured a peak-to-peak
(p--p) velocity of 440 km s$^{-1}$ (without correcting for
inclination) which is much higher than typical widths measured in
other z$\sim$1 galaxies (e.g. Vogt et al.~1997).

{\bf ISOHDFS 25}: This is an interacting system at a redshift of
z=0.58.  The two galaxies are clearly separated in the ESO--EIS K-band
image  (in Table 1 we quote magnitudes, PAs and redshifts for both of
them).  We note that ISOHDFS 25  does not have an HST-WFPC2 image
since it is outside of the WFPC2 main$/$flanking fields (the ISOCAM
HDFS field was  slightly larger than the WFPC2 one).  The separation
between the two galaxies is 2.46$^{\arcsec}$.  It is worth mentioning that
ISOCAM's final resolution for the HDF maps (see Aussel et al.~ 1999 for
a discussion) is 3$^{\arcsec}$ so that  ISOHDFS 25 is at the limit of what
ISOCAM could resolve. However, no attempt was made to resolve the
mid-infrared emission from the two components. Thus, as far as the MIR
properties of ISOHDFS 25 are concerned we will treat it as a single
object.  The two counter-rotating galaxies are clearly seen in the the
sky-subtracted  2-D spectrum of ISOHDFS 25. H$_{\alpha}$ emission is
spatially extended in both  galaxies with  an extent of $\sim$ 2
arcsec (FWHM) corresponding to about 15 kpc at the distance of the
galaxies.  The H$_{\alpha}$ is tilted in both galaxies which is
attributed to ordered  rotation. The velocity profiles  are similar in
shape and have a p-p velocity of  $\sim$300 km s$^{-1}$.

{\bf ISOHDFS 55}:  This is an Sbc type spiral at z=0.76.  The
H$_{\alpha}$ emission is clearly extended  although no continuum has
been detected. The spatial extent of the H$_{\alpha}$ line  over
$\sim$3 arcsec (smaller than the optical extent of the galaxy) implies
that the H$_{\alpha}$ does not originate in centrally concentrated gas
but most likely in a rotating disk (similar to ISOHDFS 27).   The
double-horn velocity profile gives additional evidence for the
emission originating in a ring or disk-like structure. We see no
nuclear H$_{\alpha}$ emission.  This could be due either to excess
extinction in the galaxy's nucleus or  to the presence of a
significant older stellar population component (Moy et al.~  2002, in
preparation).

\section{Dynamical Mass and the Mass-to-Light Ratio}

The dynamical mass of spiral galaxies can be estimated based on the
rotational velocity as a function of radius in the galactic
plane. Spirals are assumed to consist of two components a flat disk
and a spherical or spheroidal component  which could be a central
concentration, or a massive halo (e.g.~Bahcall~1982).  Since the
gravitational potential is a linear  function of mass, it is clear
that the actual M(R) is intermediate  between what would be predicted
for a purely flat and a purely  spherical model for the distribution
of mass and the observed rotational velocity.  The dynamical mass can
be estimated from:
\begin{equation}
M(R) = \frac{f V^{2}_{obs} R}{G sin^{2}i}
\end{equation}
where G is the gravitational constant, i is the inclination of the
system, R is the radius within which the mass is estimated and f is a
constant which has a value between 0.5 (for a disc with a flat RC) and
1.0 (for spherical distribution) independent of the presence of a
massive halo (Lequeux~1983).   The formula  is valid for any galaxy
(in equilibrium) supported by rotation.  For a disc with a flat
rotation curve f=0.6 while for a spherical distribution  (e.g. a
galaxy dominated by a dark halo) f=1.0. For a disc with  a Keplerian
decreasing rotation curve outside R, f=0.5.  Thus, according to this
model, for any rotating galaxy f should lie in the range: f=0.5 to
1.0, independent of the  presence of a massive halo.  Of course, the
mass  estimate provided by Eq. (1), will only be a good approximation
of the  true dynamical mass if the galaxy is supported by rotation. If
on the other hand the  galaxy is mainly supported by random motions,
the estimated mass  will be a severe underestimate.  For the present
mass estimates we have assumed  the most realistic spherical
distribution, i.e. f=1.0.  In addition to the intrinsic uncertainties
of this model, mass estimates of  high-redshift objects suffer from
additional uncertainties such as  estimating the extent of the object,
the inclination etc.

In Table 3 we list masses for the four ISOCAM galaxies.  With a mass
of 10$^{12}$M$_{\odot}$ (within 20 kpc) ISOHDFS 27 is  clearly a very
massive galaxy.  Its mass is higher than that of UGC 12591  (the most
massive local galaxy, (Giovanelli et al.~1986), that of L451 (a massive
disk galaxy at z=1.34, van Dokkum \& Stanford~2001) and the masses of
other high-z galaxies from the studies of Vogt et al.~(1996, 1997),
Moorwood et al.~(2000) and Pettini et al.~(2001).  ISOHDFS 27
together with L451 are examples of well formed massive disk  systems
at z$>$0.5.  The mass of ISOHDFS 27 corresponds to about seven times
the mass of an  early type galaxy at the knee of the luminosity
function (Loveday~2000, Cole et al.~2001) and this is
comparable to giant ellipticals and hosts of luminous QSOs and radio
galaxies.

In Figure 5 we investigate the correlation between dynamical mass and
bolometric luminosity for the ISOHDFS galaxies presented here, and
compare it  to that for local spirals (M33, M51), luminous and
ultra-luminous IR galaxies (LIRGS, ULIRGs, NGC 6240, NGC 3256, Arp
220) and the  prototypical starburst  galaxy M82. We also plot
M$/$L$_{bol}$ values for spirals and elliptical  galaxies. We list the
mass-to-light ratio for the ISOHDFS galaxies in Table 3.  The
bolometric luminosities of the ISOHDFS galaxies have been extrapolated
from their MIR luminosities assuming L$_{FIR}/$L$_{MIR}\sim$10
(Roussel et al.~2001) and, assuming L$_{FIR}\sim$L$_{bol}$. The dynamical
masses for the  ISOHDFS galaxies can be found in Table 3, those
for the remaining galaxies have been taken from the literature.  We
note that for the purpose of the plot in Figure 5 ISOHDFS 25 is
treated as a single object (see explanation in section 4). The ratio
M$/$L$_{bol}$ has been estimated assuming the mass of {\it both}
components, for the L$_{bol}$ the L$_{MIR}$ has been used.  To
estimate the M$/$L$_{bol}$ value for ellipticals galaxies  we used the
Bruzual \& Charlot models (Bruzual \& Charlot 1993) to compute the
fraction L$_{B} / $ L$_{bol}$, assuming M$/$L$_{B}$ $\sim$16 (e.g
Schweizer et al.~ 1989). We find that L$_{B} /$ L$_{bol} \sim$ 0.4
therefore, M $/$ L$_{bol}$ = 6.5. The value is in agreement with the
M$/$L$_{bol} \sim$ 5 measured by Arnaud \& Gilmore (1986). For the
spirals we used the M$/$L$_{bol}\sim$3.0 (e.g. Burbridge \& Burbridge
1975).

The mean M$/$L$_{bol}$ ratio for ISOHDFS 25 and ISOHDFS 55 is 0.5,
close to the value found for the IR luminous starbursting galaxy NGC
3256 (0.3).  ISOHDFS 27 on the other hand, has an M$/$L$_{bol}$ ratio
of 2.4, similar to the value for M51.  It is evident from Figure 5
that, the M$/$L$_{bol}$ ratio  is higher for earlier Hubble
types. Starburst activity however, influences the M$/$L$_{bol}$ ratio,
since bursts increase the luminosity resulting in lower M$/$L$_{bol}$
ratios.  Since the three galaxies are representative examples of the
entire ISOCAM population we conclude that the M$/$L$_{bol}$ ratio  of
the majority of ISOCAM galaxies is not as low as in local infrared
luminous starburst galaxies but rather between the values for massive
IR luminous starburst galaxies akin  to NGC 3256 and large normal
spirals.  ISOHDFS 27 represents the small fraction of ISOCAM galaxies
($\sim$ 10\%) that are large early type spirals whose large MIR
luminosity is mostly due to their large masses.

\section{Tully-Fisher Relation}

In order to look for kinematic evidence of luminosity evolution in
high  redshift galaxies one can compare the kinematical properties of
high-z galaxies to those of local galaxies.  Tully \& Fisher (1977)
showed that there exists a good correlation between  HI line width and
galaxy luminosity, the so-called Tully-Fisher (TF) relation.  Although
the original TF was established using velocity widths measured in HI
which sample the full galactic disks, Simard \& Pritchet (1998) and
Courteau (1997)  showed that there is a tight correlation between
measured HI and H$_{\alpha}$ rotation velocities so that internal
kinematics  measured from the H$_{\alpha}$ should also follow the same
TF as the HI.

In Figure 6 we compare the present data of the ISOCAM galaxies with
the local rest-frame B-band Tully-Fisher (TF) relation of
Pierce \& Tully~(1992, PT).  The rest-frame B-band data correspond
approximately to observed R-band at z$\sim$0.6 and I-band at
z$\sim$0.8.  The PT relation used data for 16 galaxies from the Ursa
Major Cluster to  define the calibration and six local galaxies to
define the zero-point.  The relation is described by:

M$_{B}$ = -7.48 (log W$_{R}$ - 2.50) - 19.55

where M$_{B}$ is the total B-magnitude and W$_{R}$ is the width of the
HI line corrected for inclination and turbulence following the
prescriptions of Tully \& Fouqu\'e (1985).  In Figure 6 we plot the
inverse TF fit i.e., V$_{rot}$ as  a function of absolute magnitude
M$_{B}$.  We have converted the PT observed linewidths W$_{R}$ to
rotational velocities following the method of Simard \& Pritchet
(1998).  We corrected the rotational velocities
for the effects of instrumental broadening, (1$+$z) cosmological 
redshift effect, sin {\it i} inclination and, finally for
extinction.  Since ISOCAM galaxies have been selected from deep MIR
surveys extinction plays a crucial role. We have estimated the
extinction in the optical A$_{V}$ based on measurements of the Balmer
decrement for a subsample of the entire ISOHDFS sample. In general we
find that the A$_{V}$ varies between 1.2--2.0 mag (Franceschini 
et al.~2002). To correct the present data we  have used
the ``conservative'' A$_{V}$ value of 1.  Using the data for the four
galaxies (for the purpose of the TF plot we have plotted both
components of ISOHDFS 25 using the available magnitudes and the
velocities measured from the present work) we compute a weighted
offset of  1.66$\pm$0.28 mag relative to the local B-band relation.

However, the B-band PT relation may not be the most appropriate for
comparison to the present ISOCAM sample since the B-band is more
sensitive to  star formation and extinction. Hence, we have compared
our ISOCAM data also to the I-band PT relation defined as (
from Pierce \& Tully~1992):

M$_{I}$ = -8.72 (log W$_{R}$ - 2.50) - 20.94

where M$_{I}$ is the total I-band magnitude and W$_{R}$ as before.  We
note that rest-frame I-band data correspond to observed   J-band at
z$\sim$0.6 and H-band at z$\sim$0.8. The I-band is more sensitive to
the underlying older stellar population and is less affected by
extinction.  In Figure 7 we show the comparison of the kinematical
data for ISOCAM galaxies to the I-band PT relation. We compute a weighted
offset of  0.7$\pm$0.26 mag relative to the local I-band relation.
{\bf Although our sample is very small for statistically significant
results, a clear offset from  the local TF is evident in both the
rest-frame B and I bands.}

A number of data-sets have so far appeared in the literature which,
based  almost exclusively on {\em rest-frame B-band spectroscopy},
have successfully probed  the TF relation at moderate redshifts. The
results can be grouped in two  categories: those finding evidence for
significant luminosity evolution  (e.g. Koo et al.~, 1995, Rix et al.~,
1997, Simard \& Pritchet 1998) and those finding modest (or no)
evolution  (e.g. Vogt et al.~ 1996, 1997).  More recently, Barden et
al.~(in preparation), studied the kinematics of a sample of
[OII]-bright objects at 0.7$<$z$<$1.5 (based on H$_{\alpha}$
rotational curves) and found significant luminosity evolution
increasing  with redshift.  The present sample of ISOCAM galaxies
differs from all previous ones since it is primarily a mid-IR selected
sample. For the selection of our objects we have not imposed any
criteria based on emission-line strengths, sizes,  and
inclination. However, the high star-formation rates found ubiquitously
among ISOCAM objects (Paper I) implies that our galaxies are in fact
strong H$_{\alpha}$ emitters although for the kinematical study we did
not preselect based on H$_{\alpha}$ strengths.

The differences in the luminosity evolution found among the various
groups implies that sample selection is in fact crucial. The galaxies
in the studies of Koo et al.~ (1995) sample the low-end of galaxy
masses. Most of their objects are compact bright emission line
galaxies. They find evidence for brightening up to 4 mags. The samples
of Rix et al.~ (1997) and Simard \& Pritchet (1998) sample larger
galaxies (scale-lengths of about 2 kpc) and luminosities in the range
L$^{*}$ to sub-L$^{*}$. They find evolution of about 1.5 mag. Vogt's
(1996, 1997) samples include even larger galaxies  (scale lengths $>$3
kpc) and the resulting brightening is modest of  the order of 0.4
mag. The present ISOCAM galaxies are large (scale lengths 7--9 kpc)
and show strong H$_{\alpha}$ emission.

Can these conflicting findings be somehow reconciled?  The answer is
yes and  to explain the apparent discrepancy one has to assume that
the luminosity evolution is highly dependent on the masses$/$sizes of
the galaxies.  In this scenario the low-mass($/$small size) galaxies
show substantial  luminosity evolution  as they undergo
interactions$/$merging and their kinematics are more susceptible to
local  phenomena such as supernovae winds and outflows. On the other
hand  the massive ($/$large size) galaxies  are less susceptible to
interactions  thus their luminosity shows less variation in comparison
to low-mass($/$small size) galaxies.  Another possibly crucial
parameter is the galaxy morphology (i.e. along the  Hubble
sequence). Although the current samples of high-z galaxies are not
large enough to allow morphological investigations, studies of local
samples  (e.g. Rubin et al.~, 1982, 1985) indicate that there is a
variation in the kinematical properties of galaxies according to their
morphological types. However, issues of internal absorption
corrections are indeed important as discussed by Kodaira \& Watanabe
(1988).  Finally, the emission-line strength also plays a role: higher
luminosity evolution is expected in objects which are undergoing (or
have recently undergone) episodes of star-forming activity compared to
the more quiescent ones.

How do the present results compare with previous studies?  From the
above discussion it is obvious that the only meaningful comparison is
with the samples of Vogt et al.~ (1996, 1997).
Despite the fact that both samples contain large (in size) galaxies
we measure a higher luminosity evolution than that found by Vogt et al.~
The measured offset is bound to increase if we use a higher extinction
value A$_{V}$ for the present ISOCAM galaxies. 
Taken at face value the difference in the luminosity evolution found
between the two samples seems to contradict the mass-dependent luminosity
evolution scenario. However, this is not the case: as we mentioned earlier,
emission-line bright objects show a higher luminosity evolution than the more
quiescent ones. A significant difference between the present
ISOCAM sample and the Vogt samples, is that the former has been drawn
from a parent sample of strong H$_{\alpha}$ emitters (Paper I). 
The Vogt et al.~galaxies were selected 
from a magnitude limited sample for having emission line flux
extending beyond the nuclear regions (so that the rotation curves 
could be traced out to the disk). However, this selection criterion does
not bias the sample towards emission-line bright objects. 
In fact,
the few Vogt et al.~galaxies, for which we have been able to find published 
Keck spectra in the literature (e.g. Forbes et al.~1996), 
are rather weak [OII]-emitters. 
We conclude
that selection criteria are important when studying TF at
high-redshifts and that the existence of a mass and emission-line strength 
dependent luminosity
evolution is probably the cause of the differences found by the
various groups.

\section{ISOCAM Galaxies and Implications for the IR Extragalactic 
Background}

With the availability of very deep 
source counts and upper limits on the diffuse isotropic emission at 
shorter wavelengths the Extragalactic Background Light (EBL)
has been reasonably constrained at visible, IR and
submm wavelengths. 
The contribution of known galaxies 
to the optical EBL has been calculated by integrating the emitted flux times 
the differential number counts down to the detection threshold. 
Madau \& Pozzetti~(2000) showed that the optical EBL is 
dominated by a large number of relatively faint galaxies 
at z$\simeq$0.7 with almost negligible contributions from  
the luminous Lyman Break Galaxies 
(e.g. those observed by Steidel et al.~(1996)) .

The discovery by COBE of an infrared EBL (e.g. Puget et al.~1996, 
Fixsen et al.~1998, Finkbeiner, Davis \& Schlegel 2000)
with an intensity similar to or greater than that of the optical EBL has 
generated a lot of interest in identifying the infrared galaxies 
responsible for it. The infrared EBL peaks around $\lambda \sim $ 140 $\mu$m.
It is thus interesting to investigate the relative contributions of the 
newly detected ISOCAM galaxies to the infrared EBL. 
Elbaz et al.~(2002)
carried out detailed calculations of the relative contributions of
various types of objects such as
luminous infrared starbursts, AGN, and normal galaxies to the infrared EBL.
With luminosities greater than 10$^{10}$L$_{\odot}$ at 8.5 $\mu$m 
(in the rest frame), a median redshift of 0.75, and bolometric 
luminosities between 2 and 10 $\times$10$^{11}$L$_{\odot}$, ISOCAM galaxies
are in fact LIRGs.
Based on their calculations, Elbaz et al.~(2002) find that the contribution 
of ISOCAM galaxies to the peak of the infrared
EBL can be of the order of 50\% or more, (i.e. a significant fraction) 
dependent
of course on the exact shape of the underlying spectral energy distribution.

Once we have established that ISOCAM galaxies are major contributors to the 
infrared EBL, it is interesting to investigate the properties of these galaxies
and compare them to the properties of the galaxies responsible for the 
optical EBL.
The discovery of four massive ISOCAM galaxies has prompted us to take an
in-depth look at
into the morphologies of both the present ISOHDFS galaxies
and the ISOCAM detections in HDF-N (Aussel et al.~1999, 
Moy et al.~in preparation).  
For the identification of the optical$/$near-IR counterparts of the
ISOHDFS galaxies we have used the WFPC2$/$F814W image and$/$or the EIS
K-band image (da Costa et al.~1998). Out of the ISOCAM sample of 63 galaxies
we were able to find secure identifications for up to 90\% of them 
(about 58 sources, for the remaining sources there is either 
no ground-based imaging followup and$/$or there was no clear counterpart). 
About half of the sources with optical$/$near-IR counterparts, appear 
to be interacting systems similar to ISOHDFS 25 while the remaining objects 
appear to be large disk dominated spirals like ISOHDFS 55 
(scale length of 7 kpc) or as large as 
ISOHDFS 27 (scale length 11 kpc about $\sim$10\% of ISOHDFS galaxies). 
Aussel et al.~(1999) 
reached similar conclusions for the HDF-N. 
Although we do not have estimates
of the dynamical mass for the majority of ISOHDFS galaxies, 
Franceschini et al.~(2002) presented baryonic mass estimates for a 
large fraction of ISOHDFS galaxies. Based on their estimated baryonic masses
and the dynamical mass estimates presented here, we conclude that ISOHDFS 
galaxies are indeed massive galaxies. 
Both ISOCAM samples (HDF-N \& S) 
show star formation rates of the order of 30--100 M$_{\odot} yr^{-1}$ 
as deduced
from their MIR luminosities (e.g. Franceschini et al.~2002).

Since the ISOHDFS galaxies presented here are representative of the 
morphologies and properties of the galaxies of the entire ISOCAM sample we
suggest that the infrared EBL is probably dominated by these few (in numbers),
large or interacting galaxies and  massive IR luminous galaxies,
as opposed to the smallish but more numerous faint blue
galaxies that were found to contribute to the optical EB 
(Madau \&Pozzetti~2000).
Thus the objects that contribute to the IR EB are not the standard IR 
counterparts of normal spiral and irregular galaxies but are IR luminous 
(L$>$few $\times$ 10$^{11}$ L$_{\odot}$), massive 
(M$>$ few$\times$10$^{11}$ M$_{\odot}$) galaxies with median redshift 0.7.

\section{Conclusions}

We have presented first results from internal kinematic studies of four 
intermediate redshift (z = 0.6 -- 0.8) galaxies detected by ISOCAM in the 
HDFS. The kinematics are derived from 
high resolution spatially resolved H$_{\alpha}$ velocity profiles.
The galaxies studied, two large Sbc spirals and a double pair$/$interacting
system, are representative of the morphologies of the entire ISOHDFS sample.
Our results are summarized as follows:

ISOHDFS galaxies tend to be massive galaxies with masses of a  
few x 10$^{11}$M$_{\odot}$. One of the galaxies studied, ISOHDFS 27, contains
a mass of 10$^{12}$ M$_{\odot}$ (within 20 kpc) substantially higher than
the masses measured in other intermediate redshift galaxies and, comparable
to the mass of giant ellipticals and hosts of luminous QSOs and radio galaxies.
The presence of a central AGN in this galaxy is ruled out by three 
independent pieces of evidence: the extend$/$ morphology of 
the H$_{\alpha}$ emission,
the H$_{\alpha}/$[NII] ratio of 3 (typical of starbursts) 
and the mid-infrared LW3(15$\mu$)m$/$LW2(7$\mu$m) ratio which again is typical
of values found in local luminous IR galaxies. 
Based on the MIR luminosities and assuming L$_{FIR}$=10 L$_{MIR}$
we have calculated the mass-to-light ratio M$/$L$_{bol}$ of ISOHDFS galaxies.
The M$/$L$_{bol}$ values are between the values found for local 
IR luminous star-forming galaxies and massive spirals. 

We have compared the four ISOCAM galaxies to the local B and I-band 
Tully-Fisher relations. We have measured offsets of 1.6$\pm$0.3 and 
0.7$\pm$0.3 mag from the B and I-band TF relations, respectively. 
Although our sample is
very small to draw statistical conclusions it is evident that a clear offset
from the local TF relations exist. We discuss the possible existence of
a mass-dependent luminosity evolution where small mass (size) galaxies
are more likely to undergo merging$/$interactions and thus show substantial
luminosity evolution. In addition, recent episodes of star-forming activity,
thus strong emission-line objects, are bound to show higher 
luminosity evolution
than the more quiescent ones. Overall, we find that selection effects are very
crucial to studies of the TF relation in 
intermediate$/$high redshift systems. We stress the uniqueness of our sample
(an infrared selected one), when comparing it to other studies. 

We have examined the global morphological properties of the ISOHDFS galaxies: 
40\% of ISOHDFS galaxies
are double pair$/$interacting systems (akin to ISOHDFS 25),
with the remaining being field disk-dominated spirals (similar to ISOHDFS 55).
A further $\sim$10\% of the spirals are large early--type systems like 
ISOHDFS 27. Based on the current dynamical mass estimates and estimates 
of the baryonic masses for the majority of the ISOHDFS galaxies 
we find that ISOCAM galaxies are indeed massive systems. 
Our results imply that the IR background is dominated by 
a small number of massive and IR luminous star-forming galaxies
near $<$z$> \sim$0.7.

\acknowledgments{We acknowledge support from the European Community 
RTN Network "POE" (under contract HPRN-CT-2000-00138). We appreciate 
discussions with Michael Rowan-Robinson regarding dust extinction and William
Vacca regarding modelling of stellar populations.}

\newpage

\centerline{\bf FIGURES}
\figcaption{Surface Brightness profiles}
\figcaption{HST$/$WFPC2 image--2D spectrum-- H$_{\alpha}$ velocity profile-- rotation
curve for ISOHDFS 27}
\figcaption{ESO--EIS image--2D spectrum-- H$_{\alpha}$ velocity profile-- rotation
curve for ISOHDFS 25}
\figcaption{HST$/$WFPC2 image--2D spectrum-- H$_{\alpha}$ velocity profile-- rotation
curve for ISOHDFS 55}
\figcaption{M$/$L ratio}
\figcaption{B-band Tully--Fisher}
\figcaption{I-band Tully--Fisher}

\begin{deluxetable}{c c c c c c c c c c c}
\tablecolumns{11} 
\tablewidth{0pc} 
\tablenum{1}
\footnotesize
\tablecaption{Photometric data} 
\tablehead{ 
\colhead{Obj}&
\colhead{RA}& 
\colhead{Dec}&
\colhead{U$^{1}$}&
\colhead{B$^{1}$}&
\colhead{V$^{1}$}&
\colhead{R$^{1}$}&
\colhead{I$^{1}$}&
\colhead{J$^{1}$}&
\colhead{H$^{1}$}&
\colhead{K$^{1}$}\\
\colhead{}&
\colhead{(J2000)}&
\colhead{(J2000)}&
\colhead{}&
\colhead{}&
\colhead{}&
\colhead{}&
\colhead{}&
\colhead{}&
\colhead{}&
\colhead{}\\
}
\startdata
S27&22 32 47.71&-60 33 35.3&22.2&21.9&21.00&20.1&19.5&18.9&
18.39&18.17\\
S25NE&22 32 45.73&-60 32 25.6&24.1&23.4&22.7&21.9&21.4&
20.27&19.80&19.58\\
S25SW&22 32 45.62&-60 32 26.1&24.4&23.6&23.0&22.0&21.6&
20.68&20.24&19.76\\
S55&22 32 58.01&-60 32 33.8&23.8&23.5&22.7&22.0&21.3&20.6&
20.05&19.88\\
\enddata
\tablenotetext{1}{magnitudes in the AB system from the EIS Deep survey 
(see text)}
\end{deluxetable}

\begin{deluxetable}{c c c c c c c}
\tablecolumns{7} 
\tablewidth{0pc} 
\tablenum{2}
\tablecaption{Morphological and spectroscopic data } 
\tablehead{ 
\colhead{Obj}&
\colhead{Type}&
\colhead{z}& 
\colhead{Flux($H_{\alpha})^{1}$}&
\colhead{Flux(LW3)}&
\colhead{A$_{V}^{2}$}&
\colhead{scale$^{3}$}\\ 
\colhead{}&
\colhead{}&
\colhead{}&
\colhead{$\times$10$^{16}$erg cm$^{-2}$ s$^{-1}$}&
\colhead{mJy}&
\colhead{mag}&
\colhead{kpc arcsec$^{-1}$}\\
}
\startdata
S27&Sb&0.58&3.28&0.39&1.5&5.9\\
S25NE&Pair&0.58&3.12&0.47$^{4}$&1.7&5.9\\
S25SW&Pair&0.58&2.47&--&1.7&5.9\\
S55&Sbc&0.76&2.41&0.20&1.8&7.2\\
\enddata
\tablenotetext{1}{observed values}
\tablenotetext{2}{A$_{V}$ measured from the Balmer decrement 
(Franceschini et al.~2002)}
\tablenotetext{3}{the physical scale of 1'' at the distance of the galaxy}
\tablenotetext{4}{includes both galaxies}
\end{deluxetable}

\begin{deluxetable}{c c c c c c}
\tablecolumns{6} 
\tablewidth{0pc} 
\tablenum{3}
\tablecaption{Kinematic modelling data } 
\tablehead{ 
\colhead{Obj}&
\colhead{inclination}& 
\colhead{R$_{d}$}&
\colhead{V$_{rot}$}&
\colhead{Mass(within radius)$^{1}$}&
\colhead{M$/$L$_{bol}$}\\ 
\colhead{}&
\colhead{deg}&
\colhead{kpc}&
\colhead{km s$^{-1}$}&
\colhead{$\times$10$^{11}$M$_{\odot}$}&
\colhead{M$_{\odot}/$L$_{\odot}$}\\
}
\startdata
S27&49$\pm$2&11&460$\pm$40&10 (20 kpc)&2.2\\
S25NE&65$\pm$5&--$^{2}$&260$\pm$45&2.3 (14 kpc)&0.54$^{3}$\\
S25SW&65$\pm$5&--$^{2}$&249$\pm$52&2.1 (14 kpc)&--\\
S55&52$\pm$2&7&270$\pm$75&2.1 (14 kpc)&0.50\\
\enddata
\tablenotetext{1}{mass within the extent of the rotation curve}
\tablenotetext{2}{no scale length has been estimated (object lies outside
of HDF-S WFPC2 and FF)}
\tablenotetext{3}{includes both galaxies}
\end{deluxetable}

\begin{figure}
\figurenum{1}
\centering
\plotone{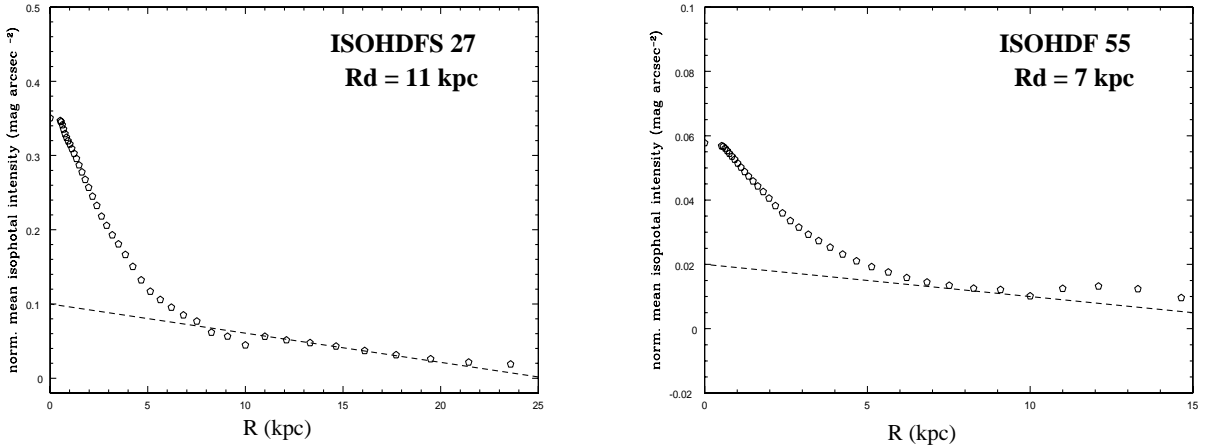}
\caption{Radial normalised intensity profiles (in magnitudes arcsec$^{-2}$)
for the two galaxies ISOHDFS 27
and ISOHDFS 55 which fall within the WFPC2 image. The profiles have been
determined with STSDAS$/$ELLIPSE. The dashed line is the exponential disk
fit.}
\end{figure}

\begin{figure}
\figurenum{2}
\centering
\caption{ISOHDFS 27, from top left clockwise: 
HST$/$WFPC2 I$_{814}$ -band 
image (North up, East to the left),
near-IR VLT--ISAAC 2D spectrum,
H$_{\alpha}$ velocity profile, and rotation curve.
The ISAAC slit width and orientation are indicated on the WFPC2 image.
The points in the velocity curve represent the observed velocities and the 
solid line is the model rotation curve.}
\end{figure}

\begin{figure}
\figurenum{3}
\centering
\caption{ISOHDFS 25, from top left clockwise:
ESO--EIS Deep K -band image,  near-IR VLT--ISAAC 2D spectrum,
H$_{\alpha}$ velocity profile, and rotation curve.
In the H$_{\alpha}$ velocity profile plot: ISOHDF S25--NE component is 
represented by a solid line, ISOHDFS 25--SW component by dotted line.
In the velocity curve plot,
filled circles and triangles represent the observed velocities  of 
the S25--NE and S25--SW component, respectively. 
Slit orientation, pixel scale and velocity curve as in Figure 2.}
\end{figure}

\begin{figure}
\figurenum{4}
\centering
\caption{ISOHDFS 55, from top left clockwise:
HST--WFPC2 I$_{814}$ -band image, near-IR VLT--ISAAC 2D spectrum,
H$_{\alpha}$ velocity profile, and rotation curve.
Slit orientation, pixel scale and velocity curve as in Figure 2.}
\end{figure}

\begin{figure}
\figurenum{5}
\centering
\includegraphics[width=12cm,angle=-90]{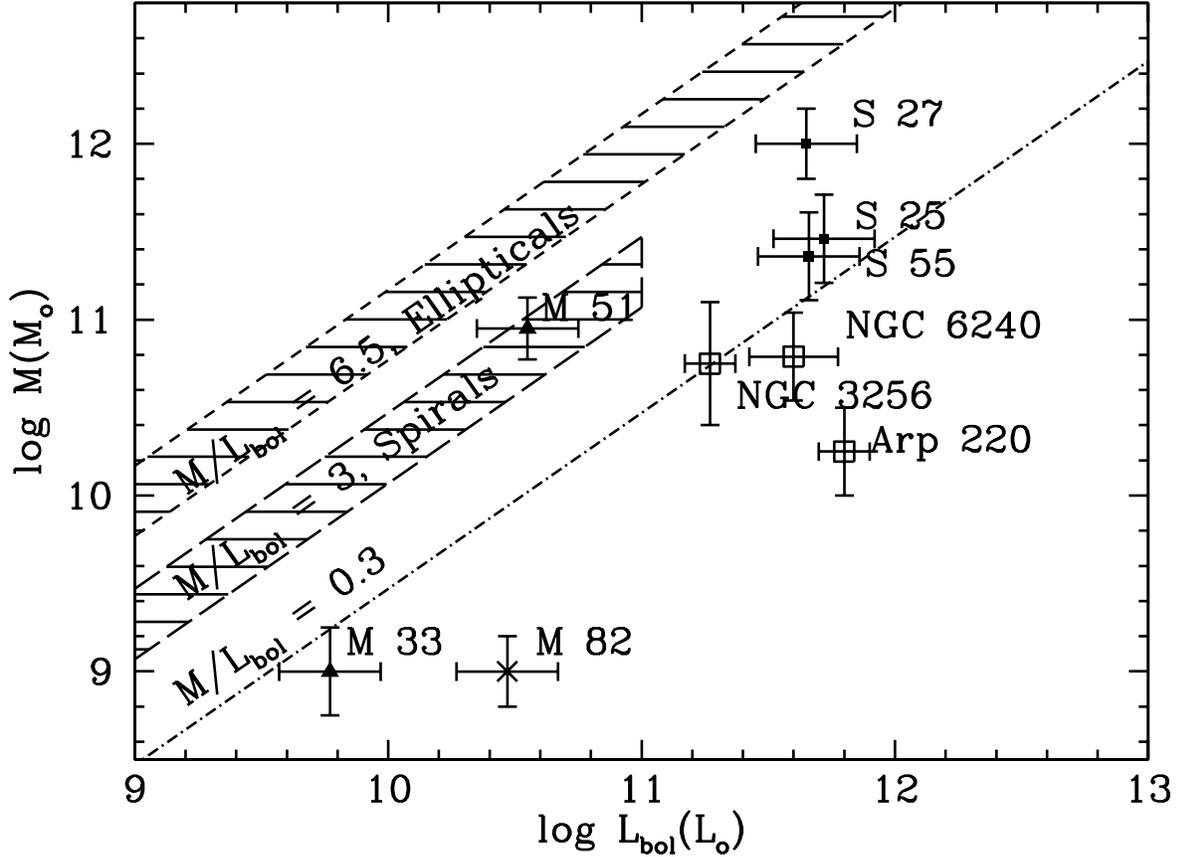}
\caption{Total masses (r$\sim$3--2 arcsec, depending on galaxy size) 
in solar units vs. bolometric luminosities for our ISOHDFS galaxies 
(filled squares), Luminous IR galaxies
(open squares), local spirals (filled triangles), starburst (cross). 
The regions occupied by Ellipticals 
(M$/$L$_{bol}$: 6.5 M$_{\odot}/$L$_{\odot}$)
Schweizer et al.~(1989) (see also text) and spirals 
(M$/$L$_{bol}$: 3 M$_{\odot}/$L$_{\odot}$)
Burbridge \& Burbrudge~(1975) are indicated. 
The NGC 3256 ratio of M$/$L$_{bol}$: 0.3 M$_{\odot}/$L$_{\odot}$ 
is also shown (dotted line). For the remaining objects shown the M$/$L$_{bol}$
values have been taken from the literature. The areas (radii) at which M
was calculated have been chosen to match the ISOCAM data. 
References: NGC 3256: Feast \& Robertson~(1978), NGC 6240:
Tecza et al.~(2000), Arp 220: Sakamoto et al.~(1999), 
M 51: Thronson \& Greenhouse~(1988),
M 33: Kormendy \& McClure~(1993), M 82: McLeod et al.~(1993).}
\end{figure}

\begin{figure}[hbtp]
\figurenum{6}
\includegraphics[width=12cm,angle=-90]{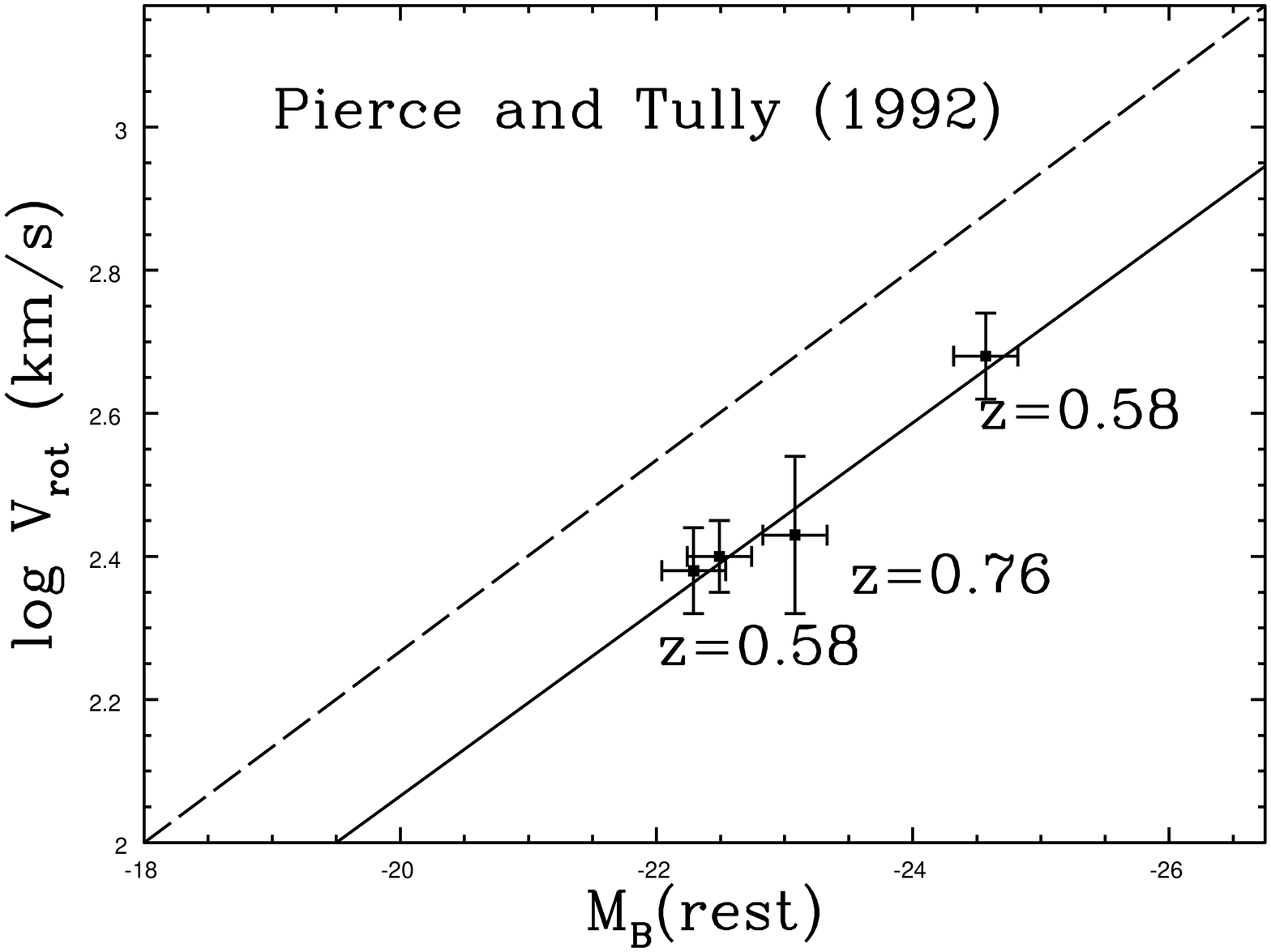}
\caption{The Tully--Fisher V$_{rot}$ as a function of 
B$_{rest}$ luminosity diagram. Our ISOCAM data are compared 
to the local TF by Pierce and
Tully (1992, dotted line). The magnitudes have been corrected for 
internal extinction and
the velocities for projection. The straight line is the fit to the ISOCAM 
points assuming same slope as the local TF. We find an offset of 
1.6$\pm$0.3 from the local TF relation.}
\end{figure}

\begin{figure}[hbtp]
\figurenum{7}
\includegraphics[width=12cm,angle=-90]{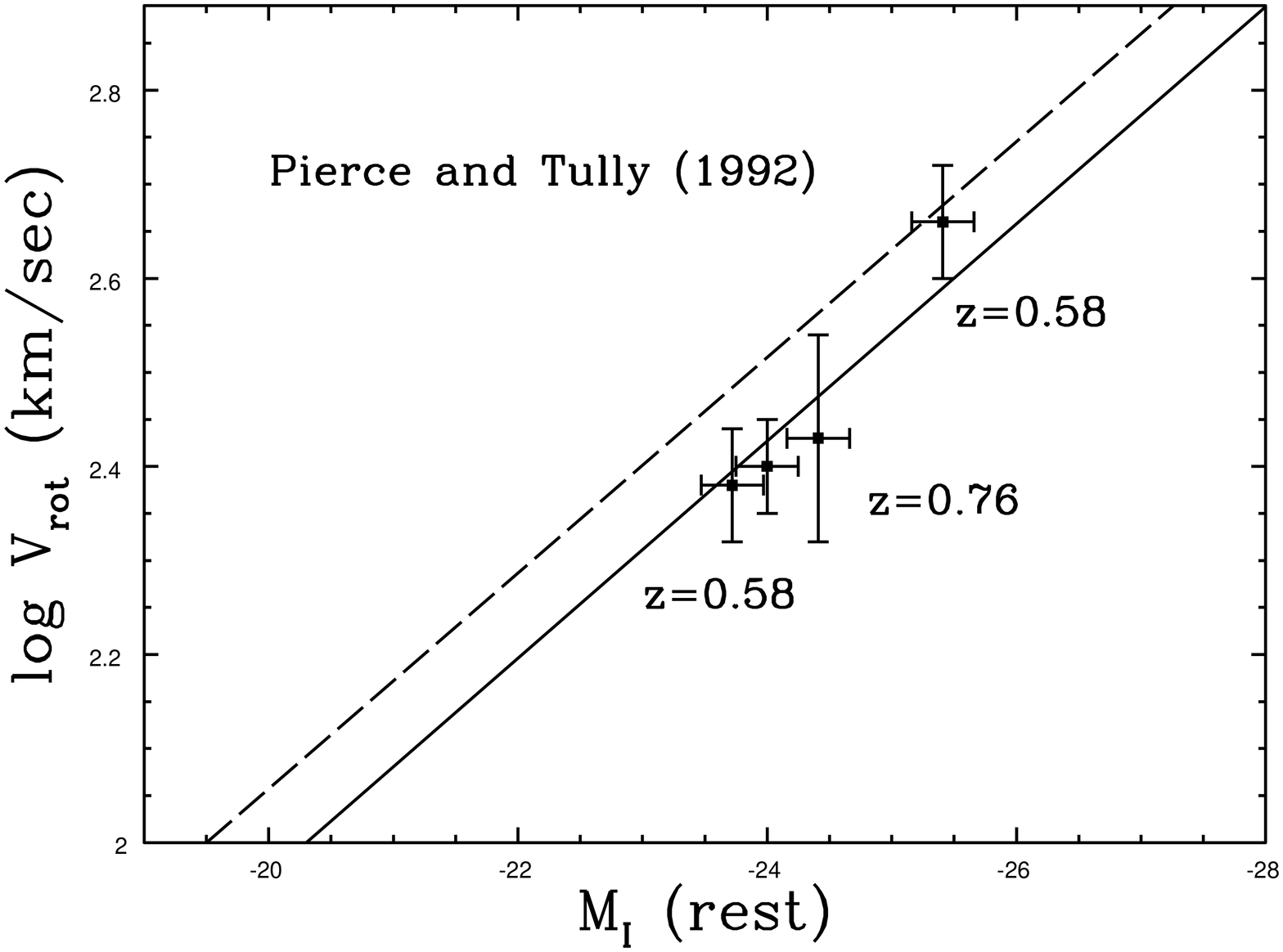}
\caption{The Tully--Fisher V$_{rot}$ as a function of 
I$_{rest}$ luminosity diagram. Our ISOCAM data are compared 
to the local TF by Pierce and
Tully (1992, dotted line). The magnitudes have been corrected for 
internal extinction and
the velocities for projection. The straight line is the fit to the ISOCAM 
points assuming same slope as the local TF. We find an offset of 
0.7$\pm$0.26 from the local TF relation.}
\end{figure}

\end{document}